\documentclass{PoS}

\usepackage{amsmath, amsfonts}
\usepackage{amssymb}
\usepackage{graphicx} 
\usepackage{subfigure} 
\usepackage{epsfig}

\def\nn{\nonumber}
\def\MS{\overline{\rm MS}}

\title{Variable Flavor Number Scheme for Final State Jets}

\ShortTitle{VFNS for Final State Jets}

\author{Andre Hoang \\
        Faculty of Physics, University of Vienna, Boltzmanngasse 5, A-1090 Wien\\
        E-mail: \email{andre.hoang@univie.ac.at}}
        
\author{\speaker{Piotr Pietrulewicz}\\
        Faculty of Physics, University of Vienna, Boltzmanngasse 5, A-1090 Wien\\
        E-mail: \email{piotr.pietrulewicz@univie.ac.at}}
        
\author{Daniel Samitz\\
        Faculty of Physics, University of Vienna, Boltzmanngasse 5, A-1090 Wien\\
        E-mail: \email{Daniel.Samitz@gmx.at}}

\abstract{We discuss a variable flavor number scheme (VFNS) for final state jets which can account for the effects of arbitrary finite quark masses in inclusive jet observables. The scheme is a generalization of the VFNS scheme for PDFs applied to setups with additional dynamical scales and relies on appropriate renormalization conditions for the matrix elements in the factorization theorem. We illustrate general properties by means of the example of deep-inelastic scattering (DIS) in the endpoint region $x\rightarrow 1$ and event shapes in the dijet limit, in particular the calculations of threshold corrections, consistency conditions and relations to mass singularities found in fixed-order massive calculations. 
}

\FullConference{XXII. International Workshop on Deep-Inelastic Scattering and Related Subjects,\\
		28 April - 2 May 2014\\
		Warsaw, Poland}

\begin{document}

\section{DIS in the classical OPE region}

The systematic treatment of massive quark effects in DIS plays an important role in the precise determination of parton distribution functions (PDFs). First we briefly summarize the crucial ingredients for a VFNS in the phenomenologically more important region $1-x\sim \mathcal{O}(1)$, before we discuss the endpoint region $x \rightarrow 1$ investigated in~\cite{Hoang2014:DIS}. We refer to~\cite{Olness:2008px} for a short overview about different implementations in literature in the OPE region.

The factorization theorem for the structure functions in DIS for massless quarks reads
\begin{align}\label{eq:fact_theorem_classical}
F_{1,2}\sim \sum\limits_{i=q,\bar{q}} e_i^2 \sum\limits_{j=q,\bar{q},g} \int_x^1 \frac{d \xi}{\xi} \,H_{ij}\left(\frac{x}{\xi},\mu\right) f_{j/P}(\xi,\mu) \, .
\end{align}
Here the sums are performed over all quark flavors with corresponding charge $e_q$. The hard matching coefficients $H_{ij}(x,\mu)$ correspond to the difference between the partonic full QCD results and corresponding low-energy expressions below the hard momentum transfer scale $Q \gg \Lambda_{\rm QCD}$ which can be described conveniently using Soft-Collinear Effective Theory (SCET)~\cite{Bauer:2002nz}. Since both theories contain the same IR behavior, the sensitivity to low-energy scales cancels in the hard matching. In the framework of SCET the PDFs $f_{i/p}(x,\mu)$ are matrix elements of operators described by collinear fields. Renormalizing them in the $\MS$ scheme, as common for massless partons, yields the DGLAP equations for the renormalization group evolution (RGE) summing the logarithms between the characteristic renormalization scale of the PDFs $\mu_f \sim \Lambda_{\rm QCD}$ and the scale of the hard interaction $\mu_H \sim Q$.

\begin{figure}
 \centering
 \includegraphics[width=7cm]{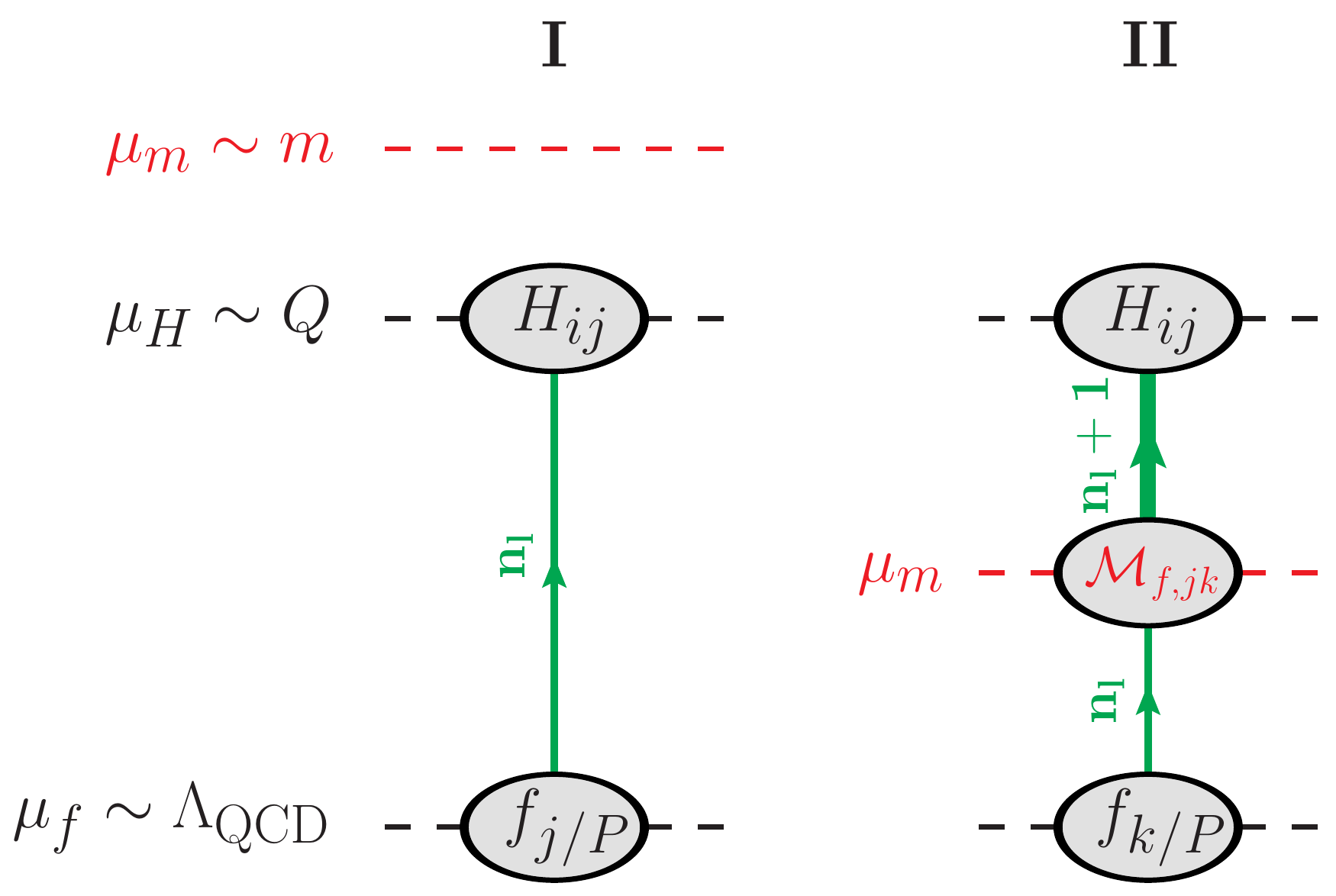}
 \caption{Illustration of a VFNS scheme for arbitrary masses in the classical OPE region $x\sim \mathcal{O}(1)$. The green arrows indicate the RG running of the PDFs and $\alpha_s$ with the corresponding appropriate number of active flavors.\label{fig:ACOT}} 
\end{figure}

In the following we consider a setup with $n_l$ massless flavors and one heavy quark with mass $m \gg \Lambda_{\rm QCD}$, which we want to incorporate in the factorization theorem in Eq.~(\ref{eq:fact_theorem_classical}).\footnote{Here we will not consider the possibility of having an intrinsic charm contribution with $m \gtrsim \Lambda_{\rm QCD}$.}
A VFNS should satisfy the following features: (i) it has to sum all large logarithms between the mass, the hard scale and $\Lambda_{\rm QCD}$, and (ii) to recover the correct limiting behavior, i.e.\ the decoupling limit for $m \gg Q$ and the massless limit for $m \ll Q$. A VFNS valid for arbitrary quark masses should continuously interpolate between these two limits (iii), also because in practice large hierarchies between the hard scale and the mass scale are rarely reached. This has been achieved in the scheme by ACOT~\cite{Aivazis:1993pi}, which we illustrate in Fig.~\ref{fig:ACOT}. The crucial ingredient is the use of proper renormalization conditions for the strong coupling and the PDFs \cite{Collins:1978wz}. Below the mass scale on-shell (OS) renormalization is employed for the virtual massive quark corrections corresponding to a low-momenutum subtraction, whereas above the mass scale one uses $\MS$ renormalization. In particular, for $m\gtrsim Q$, where always OS subtraction is used, this 
implies that the RGE for the PDFs and $\alpha_s$ is performed just with the $n_l$ massless flavors and the decoupling limit for the hard matching coefficients $H_{ij}$ is manifest for $m \gg Q$. On the other hand, for $m \lesssim Q$ the renormalization scheme is switched from OS to $\MS$ at the scale $\mu_m \sim m$. Taking into account that the quark mass does not affect the UV divergences this entails that the RGE is performed with the $n_l+1$ flavors above $\mu_m$ and with $n_l$ flavors below $\mu_m$. Due to the fact that IR mass logarithms cancel between full QCD and the effective theory description, the hard matching coefficients $H_{ij}$ smoothly approach the massless limit for $m \rightarrow 0$. The difference between OS- and $\MS$-renormalized PDFs generates a threshold correction denoted by $\mathcal{M}_{f}$ in Fig.~\ref{fig:ACOT}. 

\section{DIS in the endpoint region $x \rightarrow 1$}

\begin{figure}
 \centering
 \includegraphics[width=8cm]{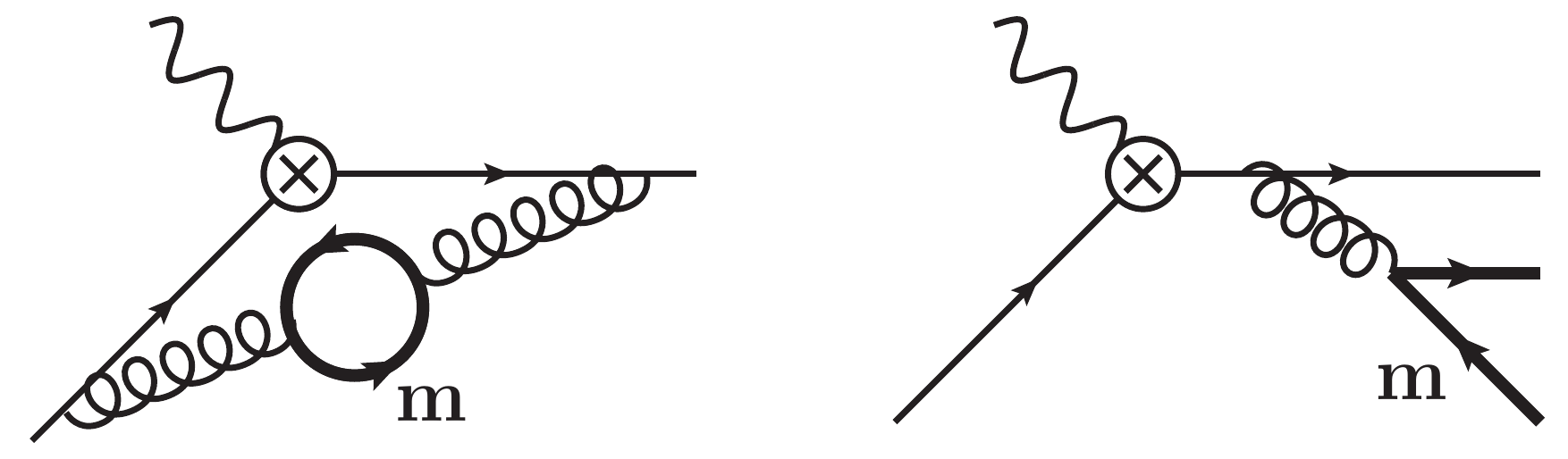}
 \caption{Examples for secondary massive quark radiation in DIS.\label{fig:DIS_secondary}} 
\end{figure}

As an instructive showcase for a VFNS with inclusive final state jets we consider DIS in the endpoint region $x \rightarrow 1$. Here the factorization theorem in Eq.~(\ref{eq:fact_theorem_classical}) is not any more appropriate due to the appearance of large logarithms $\sim \ln(1-x)$ in the hard matching coefficients $H_{ij}$. These are related to the collinear dynamics within the final state jet with a typical invariant mass of order $Q^2 (1-x) \ll Q^2$, which is a perturbative scale for $1-x \gtrsim \Lambda_{\rm QCD}/Q$. The factorization theorem for massless quarks reads in this regime up to higher orders in $1-x$~\cite{Sterman:1986aj, Manohar:2003vb, Becher:2006mr, Chay:2005rz}
\begin{align}\label{eq:fact_theorem_x1}
 F_{1,2} \sim \sum\limits_{i=q,\bar{q}} e_i^2 \,H_{\rm DIS}(Q,\mu) \int_x^1 d \xi \, J_{\rm DIS} \left(Q^2\left(\xi-x\right),\mu\right) f_{i/P}(\xi, \mu)\, .
\end{align}
Here the local hard function $H_{\rm DIS}(Q,\mu)$ is related to the matching coefficient between the full QCD and the low-energy current, and the jet function $J_{\rm DIS}(s \sim Q^2(1-x),\mu))$ describes the production rate of an inclusive jet with invariant mass $s$. The corresponding evolution factors to the common renormalization scale $\mu$ are implied. We note that no flavor-mixing terms arise in the low energy contributions to $H_{\rm DIS}$, in $J_{\rm DIS}$ and the evolution factors which leads to the fact that the parton generated out of the PDF is the one entering the hard interaction and the final state jet. The underlying reason is that the splitting of an initial collinear gluon into a collinear quark carrying the large amount of the longitudinal momentum fraction and a soft quark with momentum $\sim Q(1-x)$ is power suppressed by $\mathcal{O}(1-x)$. This feature is carried on in the presence of massive quarks, which has the consequence that the production of primary massive quarks is suppressed and massive quark effects enter mainly via secondary corrections starting at $\mathcal{O}(\alpha_s^2)$, see Fig.~\ref{fig:DIS_secondary}.\footnote{In the full QCD contributions to $H_{\rm DIS}$ there are in fact flavor non-diagonal terms starting at $\mathcal{O}(\alpha_s^3)$ for the electromagnetic vector current. Conceptually, the treatment of these terms is straightforward.}

\begin{figure}
 \centering
 \includegraphics[width=\textwidth]{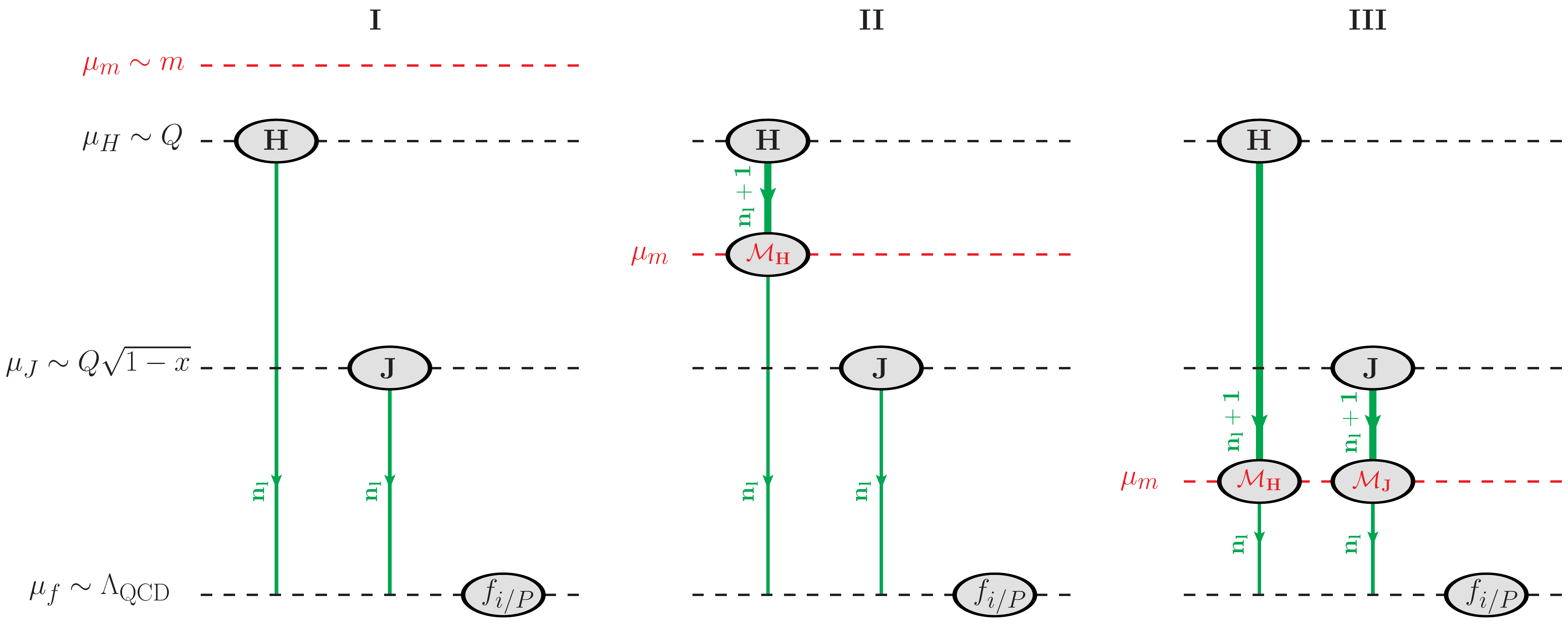}
 \caption{Illustration of the VFNS for $x \rightarrow 1$. The green arrows indicate the RG running of the hard and jet function with the corresponding appropriate number of active flavors in the top-down evolution.\label{fig:DIS2}} 
\end{figure}

Our goals for a VFNS in the endpoint region $x \rightarrow 1$ remain the same as in the classical region $x\sim \mathcal{O}(1)$ for the approach of ACOT, i.e. (i) the resummation of all large logarithms, (ii) the correct limiting behavior of the perturbative structures, i.e. $H_{\rm DIS}$ and $J_{\rm DIS}$, and (iii) a continuous description for arbitrary hierarchies w.r.~to the mass. As in the classical OPE region the use of proper renormalization conditions is crucial~\cite{Hoang2014:DIS,Gritschacher:2013pha,Pietrulewicz:2014qza}. The different possible hierarchies between the mass scale and the kinematic scales are displayed in Fig.~\ref{fig:DIS2} using top-down RGE to the PDF scale $\mu_{f}$. Note that for different values of $Q$ and $x$ the hierarchies can fall in the scenarios I, II or III. We discuss the corresponding factorization theorems schematically:

\begin{enumerate}
\item[I)]  $m \gtrsim Q$: We use OS renormalization (concerning the massive quark flavor) for the current, the jet function, the PDFs and $\alpha_s$ indicated by the superscript $(n_l)$ in the following implying an evolution with the $n_l$ massless flavors. The factorization theorem reads
  \begin{align}\label{eq:fact_theorem_I}
  F_{1,2}\sim \sum\limits_{i=q,\bar{q}} e_i^2 \, H^{(n_l)}_{\rm DIS}(Q,m,\mu_H)\, U^{(n_l)}_{H}\left(\mu_H,\mu_\Phi\right)\, J_{\rm DIS}^{(n_l)}(\mu_J) \otimes U^{(n_l)}_J(\mu_J,\mu_f) \otimes f^{(n_l)}_{i/P}(\mu_f)
  \end{align}
  The only dependence on the massive quark is located in the hard function $H^{(n_l)}_{\rm DIS}$. It contains just full QCD contributions, since the low-energy contributions vanish in the OS scheme. The massive quark decouples for $m \gg Q$. For $m\rightarrow 0$ the factorization theorem contains mass-singularities, which is, however, not the appropriate limit to be taken in this regime.
  
\item[II)] $Q \gtrsim m \gtrsim Q\sqrt{1-x}$: We use $\MS$ renormalization for the current and $\alpha_s$ above the massive threshold scale $\mu_m$ indicated by the superscript $(n_l+1)$ implying an evolution with $n_l+1$ flavors. The factorization theorem reads
  \begin{align}\label{eq:fact_theorem_II}
  F_{1,2} \sim & \, \sum\limits_{i=q,\bar{q}} e_i^2  \, H_{\rm DIS}^{(n_l+1)}(Q,m,\mu_H) \, U^{(n_l+1)}_{H}\left(\mu_H,\mu_m\right) \, \mathcal{M}_H(Q,m,\mu_m) \, U^{(n_l)}_{H}\left(\mu_m,\mu_f\right) \nn \\
  & \times J_{\rm DIS}^{(n_l)}(\mu_J) \otimes U^{(n_l)}_J(\mu_J,\mu_f) \otimes f_{i/P}^{(n_l)}(\mu_f)
  \end{align}
  The hard function $H_{\rm DIS}^{(n_l+1)}$ contains additional mass-singular (but finite) subtractions compared to $H_{\rm DIS}^{(n_l)}$ in Eq.~(\ref{eq:fact_theorem_I}), in particular also due to the now non-vanishing low-energy (SCET) current diagrams. These render $H_{\rm DIS}^{(n_l+1)}$ IR finite and yield the correct massless limit for $m \ll Q$. Below $\mu_m$ we switch to OS renormalization, which results in a massive threshold correction $\mathcal{M}_H(\mu_m)$, and the evolution is performed with $n_l$ flavors.
  
\item[III)] $Q\sqrt{1-x} > m$: We use $\MS$ renormalization for the current, the jet function and $\alpha_s$ above $\mu_m$. The factorization theorem reads
  \begin{align}\label{eq:fact_theorem_III}
   F_{1,2} \sim & \, \sum\limits_{i=q,\bar{q}} e_i^2  \, H_{\rm DIS}^{(n_l+1)}(Q,m,\mu_H) \, U^{(n_l+1)}_{H}\left(\mu_H,\mu_m\right) \, \mathcal{M}_H(Q,m,\mu_m) \, U^{(n_l)}_{H}\left(\mu_m,\mu_f\right) \nn \\
  & \times J_{\rm DIS}^{(n_l+1)}(m,\mu_J) \otimes U^{(n_l+1)}_J(\mu_J,\mu_m) \otimes \mathcal{M}_J(m,\mu_m) \otimes U^{(n_l)}_J(\mu_m,\mu_f) \otimes f_{i/P}^{(n_l)}(\mu_f) \, .
  \end{align}
  The hard function, its evolution and its massive threshold correction remain unchanged compared to Eq.~(\ref{eq:fact_theorem_II}). Now we additionally get massive quark contributions to the jet function $J_{\rm DIS}^{(n_l+1)}(m,\mu_J)$. These enter both via virtual contributions as well as via real radiation corrections for $s > 4m^2$ (see also~\cite{Pietrulewicz:2014qza}). Together they yield the known massless limit for $J_{\rm DIS}^{(n_l+1)}(m,\mu_J)$ for $m \ll Q\sqrt{1-x}$. Below $\mu_m$ we switch to OS renormalization, which results in a massive threshold correction $\mathcal{M}_J(m,\mu_m)$, and the evolution is performed just with $n_l$ flavors.
  
\end{enumerate}

Note that the massive threshold corrections $\mathcal{M}_H$ and $\mathcal{M}_J$ appearing in the factorization theorems in Eqs.~(\ref{eq:fact_theorem_II}) and~(\ref{eq:fact_theorem_III}) are directly related to the hard and the jet functions. They compensate exactly for the difference in the employed renormalization schemes and render the transitions between the factorization theorems in Eqs.~(\ref{eq:fact_theorem_I}),~(\ref{eq:fact_theorem_II}) and~(\ref{eq:fact_theorem_III}) continuous.

\begin{figure}
  \begin{center}
  \subfigure{\epsfig{file=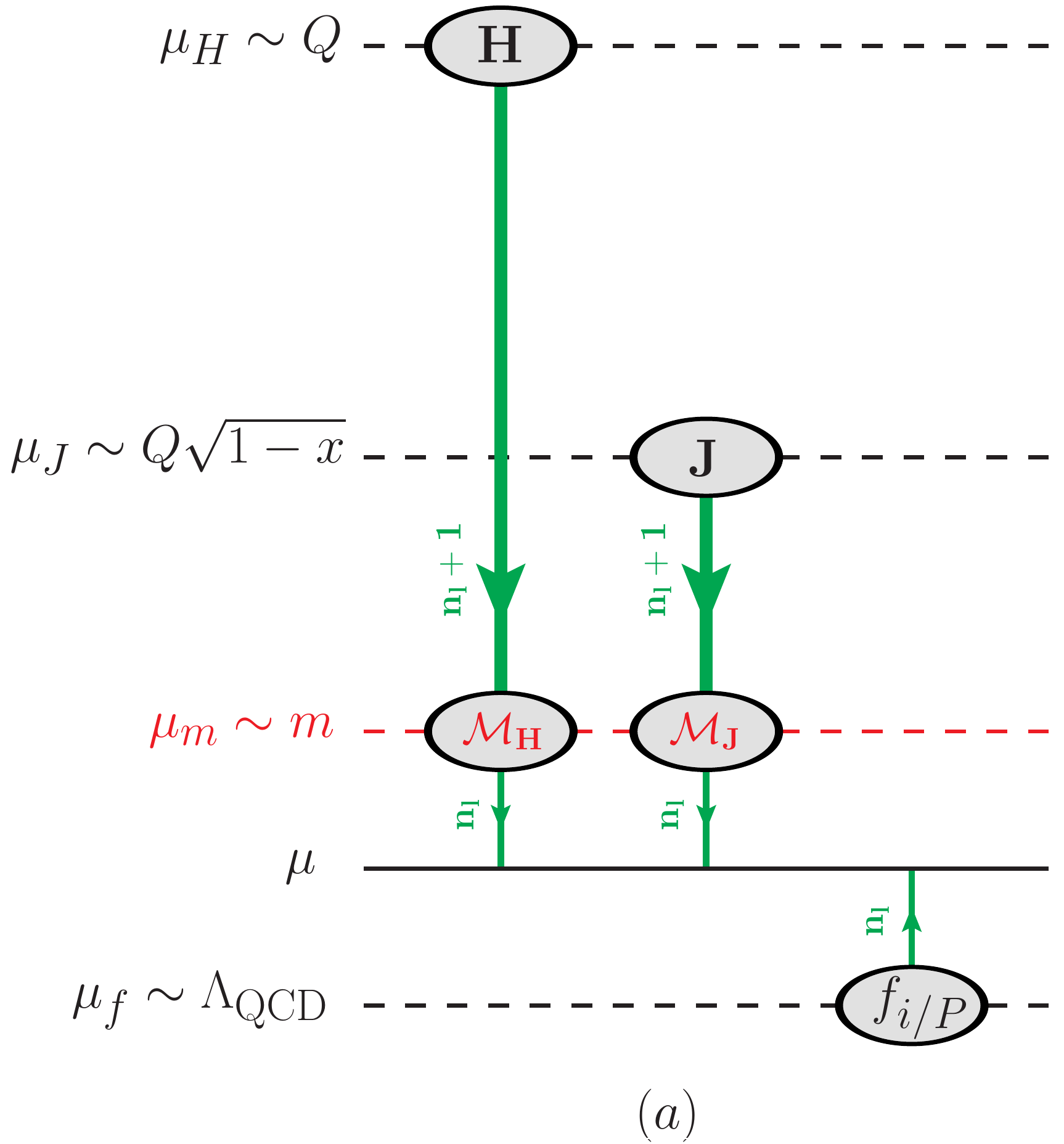,height=6 cm,clip=}}
  \subfigure{\epsfig{file=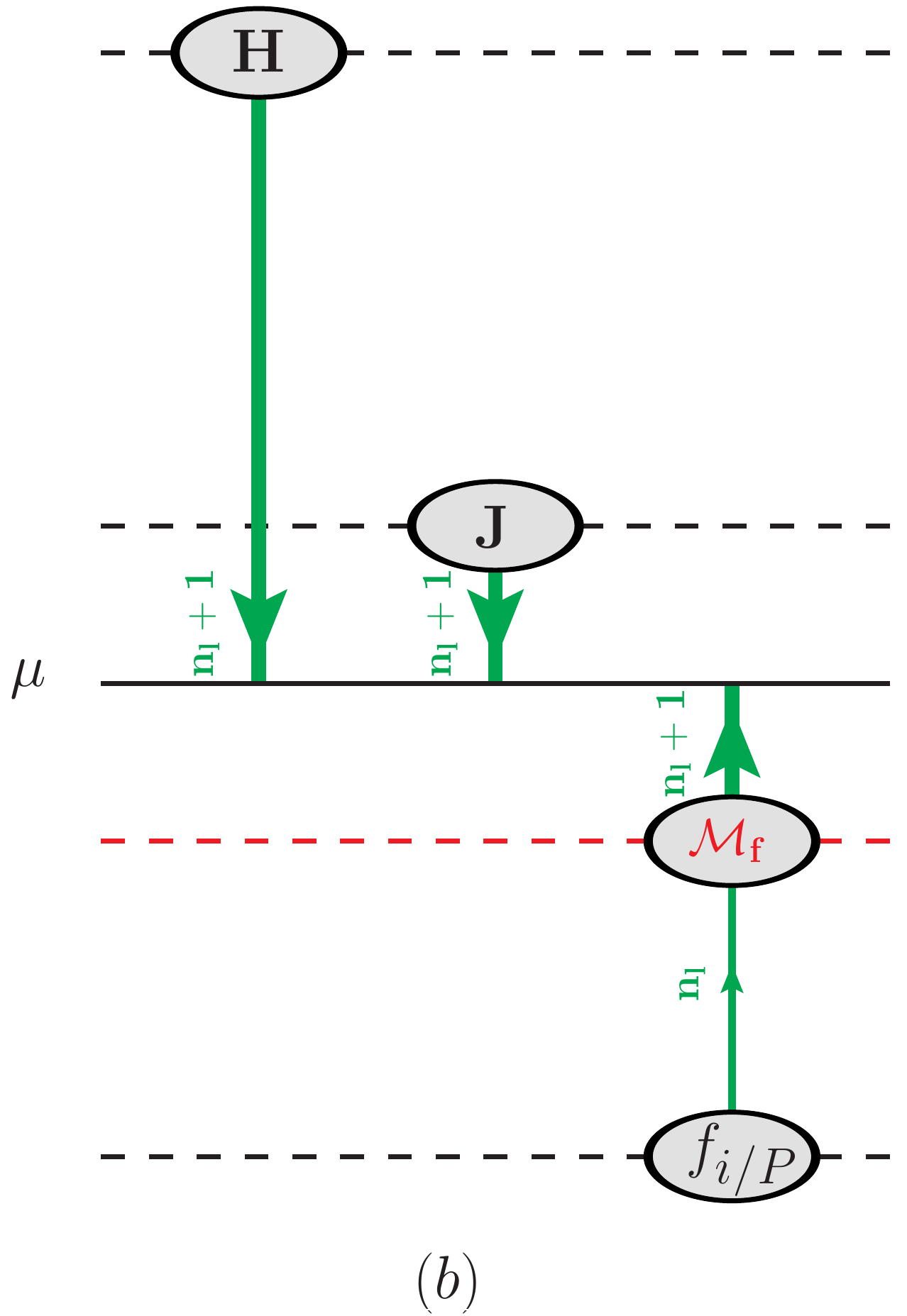,height=6 cm,clip=}}
  \end{center}
  \caption{Two different setups for RG running with $\mu<\mu_m$ (a) and $\mu>\mu_m$ (b).\label{fig:DIS_consistency}} 
\end{figure}

So far we have employed a RG setup where the hard and jet function are evolved to the generic scale of the PDF. The physical cross section is, however, independent of the final renormalization scale $\mu$. In particular we can set $\mu$ below or above the mass threshold scale $\mu_m$ as displayed in Fig.~\ref{fig:DIS_consistency} for $Q\sqrt{1-x} >m$. In the latter case the PDF is first evolved up with $n_l$ flavors, crosses the mass matching scale resulting in a threshold correction $\mathcal{M}_f$ corresponding to the change of the employed scheme, and continues its evolution with $n_l+1$ flavors. This is in analogy to the situation in the OPE region in Fig.~\ref{fig:ACOT}, where here $\mathcal{M}_f$ contains just the leading order contributions for $x\rightarrow 1$ . The equivalence to the factorization theorem in Eq.~(\ref{eq:fact_theorem_III}) implies on the one hand a relation between the evolution factors, which can be already obtained in the massless case, but on the other hand also a relation between the massive threshold corrections, namely
\begin{align}
 \mathcal{M}_H (m,\mu) \times \mathcal{M}_J (m,\mu) = \mathcal{M}_f (m,\mu) \, ,
\end{align}
which can be verified explicitly. This relation shows that the virtual contributions within the structures of the factorization theorem, i.e. the hard and jet function and the PDF, are tied together via consistency of RG running. Using a recent result for the nonsinglet heavy flavor PDF matching in Ref.~\cite{Ablinger:2014vwa} one can determine the universal threshold corrections up to $\mathcal{O}(\alpha_s^2)$ in the logarithmic counting $\alpha_s \ln(1-x) \sim 1$.

\section{Event shapes in the dijet region}

We finish with a brief discussion of event shapes in $e^+ e^-$ collisions, in particular we concentrate on thrust which we define by 
\begin{align}
\tau=1-T \,=\, 1- \frac{ \sum_i|\vec{n} \cdot \vec{p}_i|}{\sum_j |E_j|} \,=\, 1-\sum_i \frac{ |\vec{n} \cdot \vec{p}_i|}{Q}\,.
\end{align}
Here $\vec{n}$ is the thrust axis, and the sum is performed over all final state particles with momenta $\vec{p}_i$ and energies $E_i$. In the dijet limit corresponding to $\tau \rightarrow 0$ the factorization theorem for massless quarks reads to leading order~\cite{Korchemsky:1999kt,Fleming:2007qr}
\begin{align}
 \frac{\mathrm{d}\sigma}{\mathrm{d}\tau} \sim H_\tau(Q,\mu) \, \int\rm{d}\ell \, J_\tau(Q^2\tau-Q\ell,\mu) \, S_\tau(\ell,\mu) \, .
\end{align}
The hard function $H_\tau(Q,\mu)$ corresponding to the difference between the full QCD and the low-energy current and the jet function $J_\tau(s \sim \rm{max}\{Q^2\tau,Q\Lambda_{\rm QCD}\},\mu)$ describing the collinear dynamics of the two outgoing jets are analogous to DIS with corresponding replacements, i.e. $H_{\tau}= H_{\rm DIS}(Q^2\rightarrow -Q^2)$ and $J_{\tau} =  J_{\rm DIS} \otimes J_{\rm DIS}$. The main difference concerns the soft physics, where $S_\tau(\ell\sim\rm{max}\{Q\tau,\Lambda_{\rm QCD}\},\mu)$ describes now soft final state radiation between the outgoing jets which can happen at a perturbative scale in the tail region $\Lambda_{\rm QCD}/Q \ll \tau \ll 1$. The differential cross section is large in the dijet regime, which makes it phenomenologically important.

Compared to DIS a VFNS for $\tau \rightarrow0$ can be set up in an analogous way due to the similar structure of the factorization theorem. The main difference is that an addional hierarchy can arise, namely that the mass scale is below the soft scale $\mu_S \sim Q\tau$ in the tail region. In this situation we can use $\MS$ renormalization for all structures in the factorization theorem and the RG evolution does not cross the mass scale, so that no threshold corrections arise. The soft function acquires massive contributions calculated in~\cite{Gritschacher:2013tza}, which converge to the correct massless limit for $m \ll Q\tau$. A detailed discussion including a numerical analysis can be found in~\cite{Pietrulewicz:2014qza}.

\end{document}